\documentclass[a4paper,showpacs,twocolumn,amsmath,amssymb,floatfix,prb,preprintnumbers,footinbib]{revtex4}
\usepackage[dvips]{graphicx}
\usepackage{dcolumn}
\usepackage{bm}

\begin{document}


\title{Zeeman ratchets for ballistic spin currents}

\author{Matthias Scheid}
\email{Matthias.Scheid@physik.uni-regensburg.de}
\author{Michael Wimmer}
\author{Dario Bercioux}
\author{Klaus Richter}
\affiliation{Institut f\"ur Theoretische Physik, Universit\"at Regensburg, 93040 Regensburg, Germany}

\begin{abstract}
We investigate the possiblity of creating directed spin-polarized currents in a two-dimensional electron gas (2DEG) subject to an asymmetric magnetic field and an external adiabatic driving. We thereby generalize concepts of quantum charge ratchets to the case with spin. Due to the Zeeman term in the Hamiltonian, spin-up and -down electrons experience different effective potentials, which can be tailored to achieve net spin currents without corresponding charge currents. We consider ballistic, coherent transport in waveguides defined in a 2DEG, where the magnetic field modulation is induced by ferromagnetic stripes on top of the 2DEG.
\end{abstract}

\maketitle

\section{Introduction}
Magneto- and spin-electronics~\cite{zutic:2004} represent  prominent fields of modern condensed matter physics. The vision is to create electrical and opto-electronic devices with new functionalities based on spin~\cite{wolf:2001}.
Since semiconductor-based spintronics requires the generation of spin-polarized currents, various proposals have been made to this end, ranging from spin injection from a ferromagnet to adiabatic spin pumping~\cite{pumping}.

Here we propose an alternative way to generate spin currents, namely by generalizing ratchet mechanisms for creating directed particle currents~\cite{reimann:2002} to the case of spin. Particle ratchets are based on spatially periodic potentials with broken inversion symmetry which - upon ac-driving with zero net bias - give rise to a directed particle flow. They have been realized in ballistic microstructures~\cite{linke:2002}. In the present Paper we suggest to tailor spatially-periodic, non-uniform magnetic fields such that spin-up and spin-down particles exhibit opposite directed motions due to the corresponding Zeeman potential, resulting in a net spin current. A corresponding proposal for a spin ratchet based on spin-orbit interaction has been recently worked out~\cite{pfund:2006}.

\section{Model and formalism}

We consider a two-dimensional electron gas (2DEG)  in the ($x,y$) plane and a quantum wire in the $x$-direction connected to two non magnetic leads. The system is embedded in a non-uniform magnetic field $\mathbf{B}(x)=B_y(x)\mathbf{y}$ pointing in the $y$-direction. The Hamiltonian of the system reads
%
%
\begin{equation}\label{Hsim}
\mathcal{H}=\frac{p_x^2+p_y^2}{2m^*}+ \frac{g^*\mu_\text{B}}{2} B_y(x)\sigma _y+V(y),
\end{equation}
%
%
where $g^*$ is the effective gyroscopic factor, $m^*$ the effective mass, $\mu_\text{B}$ the Bohr magneton and $\sigma_y$ the Pauli spin operator. $V(y)$ denotes  the lateral confining potential. 

The particular configuration of the magnetic field permits to neglect orbital effects and allows for solving the time-independent Schr\" odinger equation through the separation ansatz: $\Phi_{\sigma}(x,y)=\varphi _{\sigma }(x) 
\chi _n(y) | \sigma \rangle_y$. Here $\chi _n(y)$ are the transversal eigenfunctions, $\varphi_\sigma (x)$ are plane waves and               $|\sigma \rangle_y$ are spin eigenvectors along the $y$-direction.
The specified magnetic field $\mathbf{B}(x)$ can be realized by making use of the fringe fields of 
ferromagnetic stripes patterned on top of the 2DEG and magnetized in the in-plane direction perpendicular to the quantum wire.
The magnetic field of a  stripe of length $b$ along the $y$-direction (with $b$ greater than the other two spatial dimensions) is in its central region $(y\approx 0)$, almost homogeneous in $y$-direction, $B_y\gg B_x,B_z$. Then the expression for the magnetic field in Eq.~(\ref{Hsim}) is valid if the quantum wire has a width $W\ll b$ and lies right beneath the center of the stripe $(y\approx 0)$. In this limit the other components, $B_x$ and $B_z$,  are small compared to $B_y$  and their effect to mix eigenstates  $|\sigma\rangle_y$ of the system can be neglected.

We are interested in the quantum transport properties of this system within the Landauer approach. For the Hamiltonian (\ref{Hsim})  the spin-flip transmission and reflection  probabilities are $T_{\sigma ,-\sigma}=T'_{\sigma ,-\sigma}=R_{\sigma ,-\sigma}=R'_{\sigma ,-\sigma}=0$ due to the suppressed mixing of the spin channels~\cite{note:1}. 
Furthermore, different lateral modes are decoupled and as consequence  it is sufficient to look at transmission and reflection probabilities of the lowest energy sub-band $(n=1)$ only, since the $m$-th mode will just mimic this behavior with an energy offset of $E_m-E_1$.  

In order to induce particle motion between the two leads, we consider an adiabatic square-wave driving as a rocking potential. The system is periodically switched between two rocking conditions, labeled by $\pm U_0$ ($U_0>0$), where it can be considered in a steady state. Therefore, the averaged net charge and spin currents inside the wire read
%
%
\begin{subequations}
\begin{equation}
\langle I_{\text{C}}\rangle =\frac{1}{2}\left[I_{\text{C}}(+U_0)+I_{\text{C}}(-U_0)\right]
\end{equation}
%
%
and
%
%
\begin{equation}
\langle I^{\mathbf{y}}_{\text{S}}\rangle =\frac{1}{2}\left[I^{\mathbf{y}}_{\text{S}}(+U_0)+
I^{\mathbf{y}}_{\text{S}}(-U_0)\right]
\end{equation}
\end{subequations}
%
%
respectively. 
%
%
\begin{figure*}
\begin{center}
\begin{minipage}[c]{0.45\textwidth}
\includegraphics[width=3in]{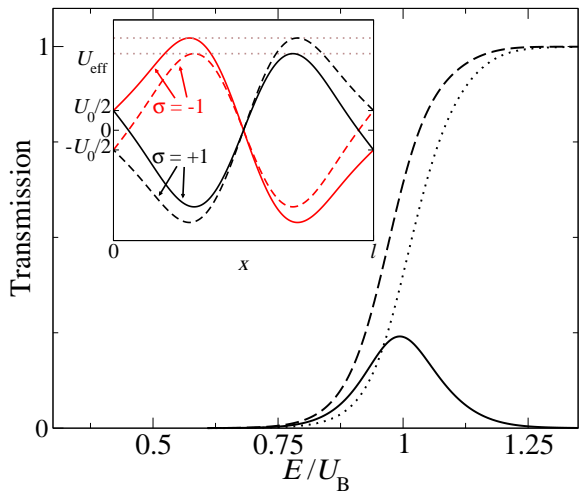}
\end{minipage}
\begin{minipage}[c]{0.45\textwidth}
\includegraphics[width=3in]{Fig2}
\end{minipage}
\caption{\label{fig:one} (left panel)
Spin-up $T_{+,+}$ (dashed line) and spin-down $T_{-,-}$ (dotted line) transmissions and their difference (solid line) as a function of the injection energy for a magnetic profile (\ref{ByTestFeld}) with $N=1$, $\alpha=0.22$ and $U_0=0.09\,U_\text{B}$ in a fixed rocking situation $+U_0$. The curves for the opposite rocking situation $-U_0$ can be obtained through the relation (\ref{Ic0Is1}). 
The Inset shows the modification of the effective Zeeman potential barrier due to the voltage drop in the two rocking situations.}
\caption{\label{fig:two}(right panel) Spin-up (dashed lines) and spin-down (solid lines) transmissions as a function of the injection energy in the two rocking situations for the magnetic profile (\ref{ByTestFeld}) with $N=1.5$, $\alpha=0.22$ and $U_0=0.15\,U_\text{B}$. 
The upper (lower) Inset shows the difference of spin-up (spin-down) transmission in the two rocking situations.}
\end{center}
\end{figure*}
%
%
\begin{widetext}
In steady state these currents are obtained within the Landauer formalism as:
%
%
\begin{subequations}\label{ave:currents}
\begin{align}
I_{\text{C}} &=-\frac{e}{h}\int_{0}^{\infty} \textrm{d}E \left[ f(E;\mu_\text{L}) - f(E;\mu _\text{R} ) \right] \left[ T_{+,+}(E) + T_{-,-}(E) \right]\,, \label{charge}\\
I^{\mathbf{y}}_{\text{S}}&=\frac{1}{4\pi}\int_{0}^{\infty}  \textrm{d}E \; \left[ f(E;\mu_\text{L} ) - f(E;\mu_\text{R})  \right] 
\left[ T_{+,+}(E) - T_{-,-}(E)\right]. \label{spin}
\end{align}
\end{subequations}
%
%
\end{widetext}
In Eqs.~(\ref{ave:currents}), $T_{+,+}(E)$ and $T_{-,-}(E)$ are the spin resolved transmissions, and $f(E;\mu_\text{L/R})$ is the Fermi function for the left/right lead at chemical potential $\mu_\text{L/R}$. 
The absence of processes that can change the spin state of the electron causes the spin current to be conserved throughout the wire.

Charge and spin currents (\ref{ave:currents}) do not vanish in the non-linear regime. To model a finite voltage drop across the two leads, we add the term $\mathcal{H}_U(t)=-U(t) g(x,y;U)$ to the Hamiltonian (\ref{Hsim}). Here the function $g(x,y;U)$ describes the spatial distribution of the electrostatic potential inside the mesoscopic system and in general is obtained through a self-consistent solution of the many-particle Schr\"odinger equation and the Poisson equation.

In the present Paper we employ a heuristic model for the voltage drop $g(x,y;U)$~\cite{linke:2002}. We assume that in the central part of the mesoscopic system the voltage drops locally proportional to  the gradient of the magnetic field $\frac{\partial}{\partial x}g(x,y;U)\propto \left|\frac{\partial}{\partial x}B_y(x) \right|$. This model is justified by the fact that the Zeeman term acts as an effective potential barrier, and it has been shown~\cite{xu:1993} that a more rapid potential variation leads to stronger wave reflection and hence to a more rapid local voltage drop. As a direct consequence, the motion of spin-up and spin-down electrons can give rise to different rectified currents, opening the possibility for the spin ratchet effect. 

%
%
\begin{figure*}[ht]
\begin{center}
\begin{minipage}[l]{0.45\textwidth}
\includegraphics[width=3in]{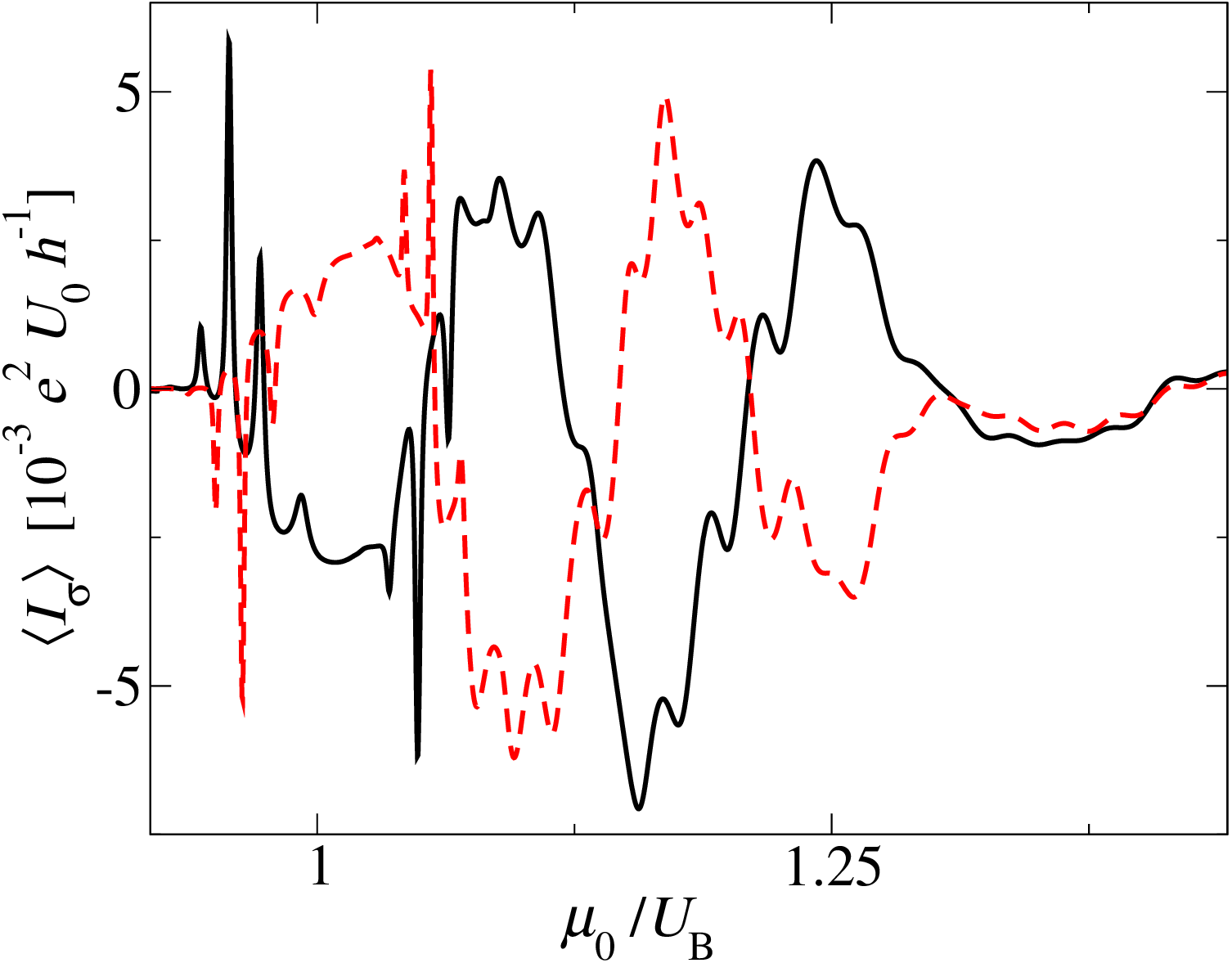}	
\end{minipage}
\begin{minipage}[r]{0.45\textwidth}
\includegraphics[width=3in]{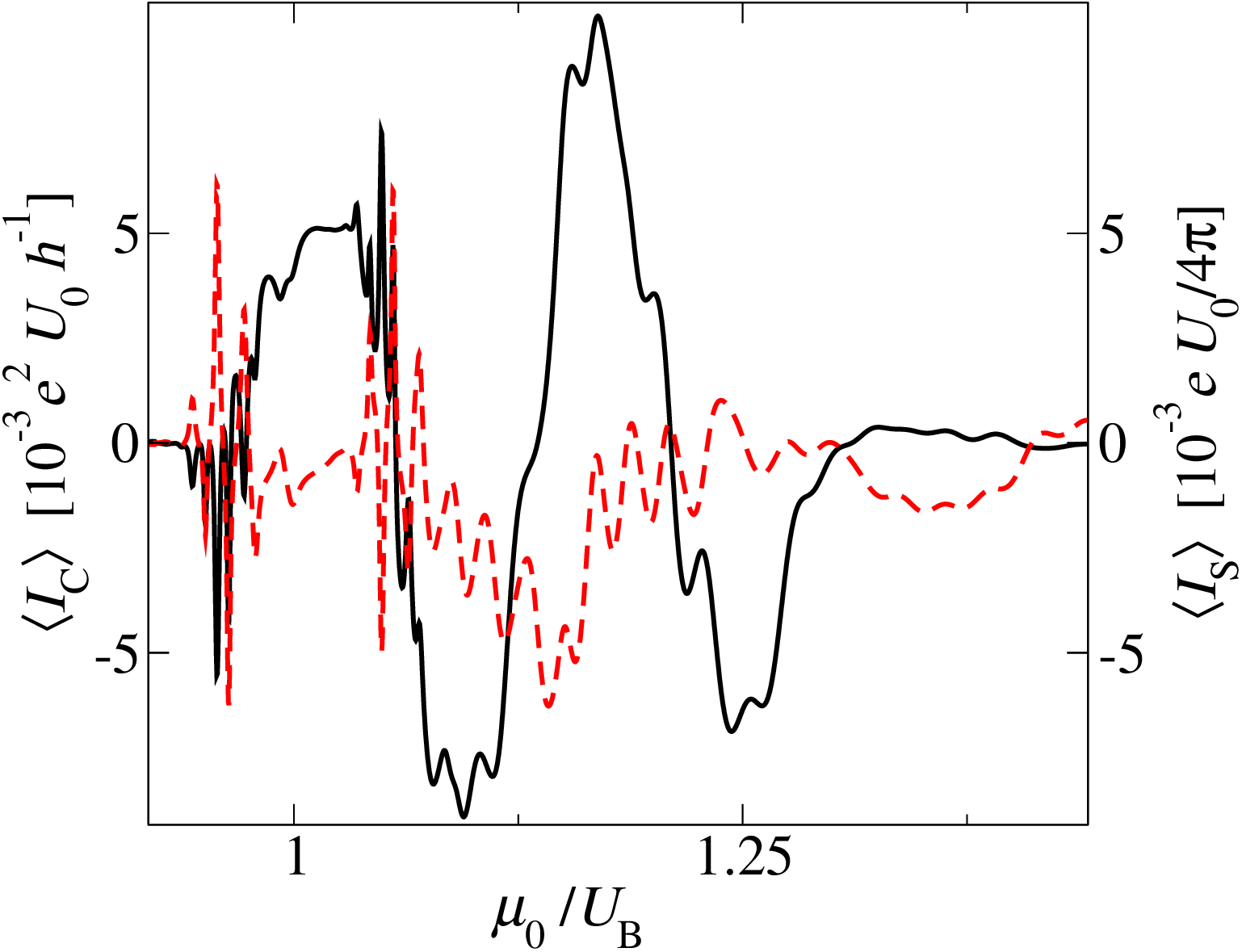}
\end{minipage}
\caption{\label{fig:three}(left panel) Averaged spin-up (solid line) and spin-down (dashed line) current as a function of the Fermi energy for a magnetic profile with $N=7.5$, $\alpha=0.22$ and $U_0=0.09\,U_\text{B}$. }
\caption{\label{fig:four}(right panel) Averaged spin (solid line) and charge (dashed line) current as a function of the Fermi energy  for a magnetic profile with $N=7.5$, $\alpha=0.22$ and $U_0=0.09\,U_\text{B}$.}
\end{center}
\end{figure*}
%
%
\section{Numerical results}
The energy dependent transmissions $T_{+,+}$ and $T_{-,-}$ are numerically calculated from lattice Green functions.
Within the scattering region of length $L$ and width $W$, an array of magnetic stripes with antiparallel magnetization can be chosen such that the resulting magnetic field is given by 
%
%
\begin{equation}
B_y(x)= B_0\left[ -\sin\left(\frac{2\pi}{l}x\right)+\alpha\sin\left(\frac{4\pi}{l}x\right) \right]\,.
\label{ByTestFeld}
\end{equation}
%
%
Here $l$ is the spatial period of the non-uniform magnetic field, and $\alpha$ is an asymmetry parameter.
The ratio $N=L/l$ is the number of magnetic field unit cells within the scatterer.
The system parameters are chosen to have the experimentally accessible  values~\cite{gui:2004}: $m^*=0.1m$,~$g^*=40$, $W=100$nm and $l=2\mu$m. The magnitude of the magnetic field $B_0$ is fixed to $0.1$T.

The system Hamiltonian (\ref{Hsim}), in presence of the magnetic field  (\ref{ByTestFeld}) and $N$ integer, shows the symmetry relations $[ \mathcal{H}, \sigma _x\hat{R} _x ] =0$ and $[ \mathcal{H}, \sigma _z\hat{R} _x ] =0$, where $\hat{R}_x$ is the operator for reflection at the $x$-axis.
Those allow us to interrelate the elements of the scattering matrix for the two rocking situations. 
We find a non-zero averaged  spin current
%
%
\begin{equation}
T_{\sigma,\sigma}(+U_0) = T_{-\sigma,-\sigma}(-U_0) \quad\Rightarrow\quad
\langle I^{\mathbf{y}}_{\text{S}}\rangle =I^{\mathbf{y}}_{\text{S}}(+U_0)
\label{Ic0Is1}
\end{equation}
%
%
and vanishing averaged charge current $\langle I_{\text{C}}\rangle =0$.

In Fig.~\ref{fig:one} we present for the case of $N=1$ the transmission probabilities as a function of the injection energy normalized to the maximum height of the magnetic barrier, $U_\text{B}=g^*\mu_\text{B}\max[B_y(x)]/2$. 
For a fixed rocking situation, spin-up and spin-down electrons pass through opposite magnetic barriers with a different maximum height due to the effective voltage drop (see Inset of Fig.~\ref{fig:one}). This causes the difference between the $T_{+,+}$ and $T_{-,-}$ transmissions.
Apart from this classical effect, this transmission difference can also have a quantum origin, \emph{e.g.} 
tunneling, relying on the asymmetric shape of the barriers ($\alpha\neq 0$).
To exclude the classical effect, we turn to a magnetic field profile where the same voltage drops from the high potential reservoir to the first barrier in either of the two rocking situations. This can be achieved by extending the size of the scattering region in (\ref{ByTestFeld}) by half a unit cell, \emph{i.e.} $N=1.5$.

Since $N$ is no longer an integer, Eq. (\ref{Ic0Is1}) does not hold true anymore. Rectification of spin-up and spin-down 
electrons still may provide us with non-zero averaged spin current, but due to the absence of symmetry 
the averaged charge current does not vanish anymore. 
In Fig.~\ref{fig:two} we show the energy dependence of the transmission probabilities  for both spin species in each rocking situation and in the Insets their difference for $N=1.5$  and $\alpha=0.22$.
The main difference between $T_{+,+}$ (black dashed lines) and $T_{-,-}$ (red solid lines) is due to the peculiar choice of the magnetic field profile: spin-up electrons travel through a single barrier whereas spin-down electrons experience a double barrier structure. Indeed, this double barrier structure gives rise to resonant tunneling behavior for $T_{-,-}$. Due to this fundamental difference it is intuitively clear that the rectification for both spin species is not opposite to each other and therefore results in finite averaged charge and spin currents. 

Increasing the number of unit cells of the magnetic profile and thereby approaching the setting of a periodic ratchet, it is possible to observe how the resonances are giving rise to the formation of minibands. 
In Fig.~\ref{fig:three}  we show the averaged spin resolved currents as a function of the Fermi energy $\mu_0$. Those are evaluated for a magnetic profile with $N=7.5$, $\alpha=0.22$ and the magnitude of the applied voltage is $U_0=0.09\, U_\text{B}$. It is evident that for almost all the values of $\mu_0$ in the plot the averaged spin-up current has opposite direction and approximatively equal magnitude with respect to the spin-down current.  

In Fig.~\ref{fig:four} we show the averaged spin and charge current as a function of the Fermi energy $\mu_0$ for the same parameters as in Fig.~\ref{fig:three}. The averaged charge current (dashed line) is mostly small compared to the averaged spin current (solid line).
The sharp peaks and the oscillating structures in Figs.~\ref{fig:three} and \ref{fig:four} can be attributed to the minibands of the magnetic superlattices. 

Those results confirm our expectation that the asymmetry in the barriers of the magnetic field can translate 
into different, sometimes even opposite rectification for spin-up and spin-down electrons.

\begin{acknowledgments} 
This research has been supported by the Deutsche Forschungsgemeinschaft within the cooperative research center SFB 689 ``Spin phenomena in low dimensions''.
\end{acknowledgments}


\begin{thebibliography}{99}

\bibitem{zutic:2004} I. \v Zut\'\i c, J. Fabian, and S. Das Sarma, Rev. Mod. Phys. {\bf 76}, 323 (2004).

\bibitem{wolf:2001} S.A. Wolf \textit{et al.}, Science \textbf{294}, 1488 (2001).

\bibitem{pumping} E.R. Mucciolo, C. Chamon, and C.M. Marcus, Phys. Rev. Lett. \textbf{89}, 146802 (2002); M. Governale, F. Taddei, and R. Fazio, Phys. Rev. B \textbf{68}, 155324 (2003).

\bibitem{reimann:2002} P. Reimann, Phys. Rev. \textbf{361}, 57  (2002).

\bibitem{linke:2002} H. Linke \emph{et al.}, Science \textbf{268}, 2314 (1999).

\bibitem{pfund:2006} A. Pfund, D. Bercioux, and K. Richter, cond-mat/0601118.

\bibitem{note:1} The channel mixing is the precondition for generating spin current  in the case of spin-orbit based ratchet~\cite{pfund:2006}.

\bibitem{xu:1993} H. Xu, Phys. Rev. B \textbf{47}, 15630 (1993).

\bibitem{gui:2004} Y. G. Gui \emph{et al.}, Europhys. Lett. \textbf{65}, 393 (2004).

\end{thebibliography}
\end{document}